\documentstyle[12pt]{article}

\begin{document}
\thispagestyle{empty}

\begin{center}
               RUSSIAN GRAVITATIONAL ASSOCIATION\\
               CENTER FOR SURFACE AND VACUUM RESEARCH\\
               DEPARTMENT OF FUNDAMENTAL INTERACTIONS AND METROLOGY\\
\end{center}
\vskip 4ex
\begin{flushright}                           RGA-CSVR-001/94\\
                                             gr-qc/9403059
\end{flushright}
\vskip 45mm
\begin{center}

{\bf ON POSSIBLE MEASUREMENT\\
     OF GRAVITATIONAL INTERACTION PARAMETERS\\
     ON BOARD A SATELLITE\\
\vskip 5mm
     A.D.Alexeev, K.A.Bronnikov, N.I.Kolosnitsyn,\\
     M.Yu.Konstantinov, V.N.Melnikov and A.G.Radynov}\\
\vskip 5mm
     {\em Centre for Surface and Vacuum Research,\\
     8 Kravchenko str., Moscow, 117331, Russia}\\
     e-mail: mel@cvsi.uucp.free.msk.su\\
\vskip 60mm

             Moscow 1994
\end{center}
\pagebreak

\setcounter{page}{1}
\begin{center}

{\bf ON POSSIBLE MEASUREMENT\\
     OF GRAVITATIONAL INTERACTION PARAMETERS\\
     ON BOARD A SATELLITE\\
\vskip 5mm
     A.D.Alexeev, K.A.Bronnikov, N.I.Kolosnitsyn,\\
     M.Yu.Konstantinov, V.N.Melnikov and A.G.Radynov}
\end{center}
\vskip 10mm
ABSTRACT
\vskip5mm

\noindent    The recently suggested SEE (Satellite Energy Exchange)
     method of measuring the gravitational constant $G$, possible
     equivalence principle violation (measured by the E\"{o}tv\"{o}s
     parameter $\eta$) and the hypothetic 5th force parameters
     $\alpha$ and $\lambda$ on board a drag-free Earth's satellite is
     discussed and further developed. Various particle trajectories
     near a heavy ball are numerically simulated. Some basic sources
     of error are analysed. The $G$ measurement procedure is modelled
     by noise insertion to a ``true'' trajectory. It is concluded
     that the present knowledge of $G, \alpha$ (for $\lambda \geq
     1$ m) and $\eta$ can be improved by at least two orders of
     magnitude.
\vskip15mm

          The gravitational constant $G$ is at present the least
     accurately measured fundamental physical constant: the error
     $\delta G/G$ is about $10^{-4}$ , while the other constants
     are known up to $10^{-6}$ or better [1-4]. Despite the repeated
     suggestions of laboratory $G$ measurements at the level of
     $10^{-5}$ not a single group has penetrated beyond $10^{-4}$;
     moreover, three of the four best absolute $G$ determinations are
     at variance with each other at their accuracy levels. There also
     exist some geophysical data on $G$ which disagree with the
     laboratory ones [2].

          Apparently suggestions to measure $G$ and other
     gravitational interaction parameters in space, by precision
     tracking the motion of artificial bodies ([5,6] and others), are more
     promising: one can avoid environmental influences difficult to
     account for and create such conditions that a particle be not
     subject to forces much greater than those under study.

          The approach of Ref.6 is to study the relative motion of two
     bodies on board a drag-free Earth's satellite using the horseshoe
     type trajectories [7]: the lighter body (``particle''), moving
     along a lower orbit that the heavier one (``shepherd''),
     overtakes it and, due to their gravitational interaction, gains
     energy, passes to a higher orbit and begins to lag behind (the
     Satellite Energy Exchange, or SEE method). The interaction phase
     can be studied within a drag-free capsule (a cylinder 20 m long,
     about 1 m in diameter) where the particle can remain as long as
     $10^{5}$ seconds.  By [6], particle trajectory measurements
     enable one to improve the existing knowledge of $G$ by 2 orders
     of magnitude.  Moreover, the 5th force parameter $\alpha$ for a
     certain range of interaction lengths $\lambda$ and the possible
     equivalence principle violation parameter (the E\"{o}tv\"{o}s parameter
     $\eta$) can be also measured with an unprecedented accuracy.
     Ref.[6] contains a number of details of the proposed experiment,
     in particular, it is shown that optimum orbital heights $H$
     range from 1390 to 3330 km.

          We have carried out a further study of the SEE method. As
     compared with [6], a wider range of particle trajectories has
     been investigated, various sources of error have been studied and
     some new estimates concerning the capabilities of the method have
     been obtained. The results are as follows.

\vskip 1.0ex
          1. The particle motion is governed by tidal and inertial
     forces and by interaction with the shepherd. Estimates of
     influence of different factors on particle motion are given in
     Table 1. The upper bounds of displacements are estimated as
     $\delta l = at^{2}/2$ assuming that an acceleration $a$ acts in
     the same direction for the time $t$ (either $10^{4}$ s, or half
     orbital period, i.e., about 1 hour, for external tidal forces
     whose influence is actually periodical). For definiteness we
     assumed that the orbital radius is $a = 8000$ km and the
     particle-shepherd distance is 10 m (half length of the capsule).
     The value of $\delta l$ is of particular interest since it is the
     particle position that is actually measured.

\vskip5mm
{\bf Table 1}

{\bf Contributions to particle motion dynamics}

\vskip5mm
\begin{tabular}{|l|c|c|}\hline

                          &     Acceleration       &     Resulting\\
         Factor           &        created         &   displacement \\
                          &      ($cm/s^{2}$)      &  for $t\sim 10^{4}$s\\
                                                              \hline

                          &                        &    \\
 1. Quadrupole tidal forces &    $\sim 10^{-8}$    &     $\sim 10$\\

 2. Higher                &                        &                \\
\hspace{5mm}geopotential harmonics& $\sim 10^{-12}$ & $\sim 10^{-4}$\\

 3. Solar tides           &$\sim 7\cdot 10^{-11}$  &  $\sim 3\cdot 10^{-4}$\\

 4. Lunar tides           &$\sim 3\cdot 10^{-10}$  &   $\sim 10^{-3}  $\\

 5. Jovian tides          &$\sim 5\cdot 10^{-16}$  &  $\sim 2\cdot 10^{-8} $\\

 6. Lunar nonsphericity   &$\sim 5\cdot 10^{-18}$  &  $\sim 2\cdot 10^{-10}$\\

 7. Relativistic tides    &  $\sim 10^{-12} $      &  $\sim 3\cdot 10^{-5} $\\

 8. Uncertainty of        &                        &                     \\
\hspace{5mm}shepherd's orbit & $\sim 3\cdot 10^{-13}$  &   $\sim 10^{-5}$ \\

 9. Possible EP violation &                        &                     \\
\hspace{5mm}($\eta = 10^{-13}$) &$\sim 7\cdot 10^{-11}$ & $\sim 3\cdot
10^{-3}$\\
                          &                        &       \\
\hline
\end{tabular}
\vskip5mm

          Assuming that the measurement error is no less than
     $10^{-6}$ cm (about 1/50 of the visible light wavelengths), the
     factors 5 and 6 from the table are manifestly negligible, like
     many others of similar origin. The factors 2,3,4,7 are to be
     included in the computer routine of an actual experiment but can
     be neglected at the planning stage aimed at working out the
     experiment strategy.

          Effects changing the satellite orbit are not included since
     the actual orbit is assumed to be known from radar or laser
     measurements. However, the corresponding (possibly systematic)
     error implies tidal acceleration uncertainties as reflected in
     line 8 of the table. One has to conclude that this uncertainty is
     a key factor for the experiment viability since a better accuracy
     than that to $\Delta R \sim 1$ cm is not expected in the coming
     years and even 1 cm is questionable. On the other hand, it makes
     no sense to measure particle positions up to a certain $\delta l$
     for such a period $t$ that the above uncertainty is greater
     than $\delta l$. For instance, if $\Delta R = 1$ cm and $\delta l =
     10^{-6}$ cm, a particle trajectory measurement should not last
     longer than  $\sim 3000$ s $\sim 1$ hour.

\vskip 1.0ex
          2. We considered the equations of particle motion with
     respect to the shepherd for arbitrary satellite orbits and
     arbitrary capsule orientations, including linear and quadratic
     terms in the ratio $s/R$ where $s$ is the shepherd-particle
     distance and $R$ is the shepherd's separation from the Earth's
     centre, which provided the required calculation accuracy. It has
     proved to be impossible to find even approximate analytic
     solutions, even for the simplest situation of particle motion in
     the plane of a circular orbit of the shepherd in the spherically
     symmetric Newtonian field of the Earth when the equations are
\begin{eqnarray}
      \ddot{x}- 2\omega\dot{y} = 3\omega^{2}xy/a + (M+m)(x/s)dU/ds \\
     \ddot{y}+ 2\omega\dot{x}=3\omega^{2}y+
         3\omega^{2}(x^{2}-2y^{2})/(2a)+ (M+m)(y/s)dU/ds.      
\end{eqnarray}
     Here $x$ is a backward along-track coordinate, $y$ is directed
     from the Earth along the geocentric radius vector and $\omega =
     (GM_{E}/a^{3})^{1/2}$ is the orbital frequency ($M_{E}$ is the
     Earth's mass). The potential $U(s)$ can include, along with the
     Newtonian term $G/s$, the 5th force potential
     $(G \alpha/s)\exp(-s/\lambda)$ or several terms of this sort.

          A possible EP violation at distances of the order of the
     Earth's radius leads to emergence of an additional term of
     the form $-\eta\omega^{2}a$ at the right-hand side of (2).

          Elliptic satellite orbits and (or) inclusion of the Earth's
     quadrupole gravitational potential lead to certain complications
     in the equations of motion.

\vskip 1.0ex
          3. In our computer simulations we solved the particle
     equations of motion for the following shepherd orbits in the
     Earth's Newtonian gravitational field:  (i) circular in
     spherical field (Eqs. (1) and (2)); (ii) circular equatorial, in
     spherical plus quadrupole field; (iii) elliptic with
     eccentricities up to 0.05 in spherical field. The rational
     extrapolation method was used, with a variable integration step
     and accuracy control. In some cases parallel calculations were
     performed by the Runge-Kutta method, by the 5th order Adams
     method and by calculations with time reversal (from the finish
     to the start of the same trajectory). It was concluded that the
     computational error was within $10^{-10}$ cm, far beyond the
     achievable measurement accuracies.

\vskip 1.0ex
          4. Part of the simulations used the so-called standard
     initial data (SID), i.e., those corresponding to particle motion
     along a nearby circular orbit, or, in case (iii), an elliptic
     one with the same eccentricity.

          Typical families of trajectories for the case (i) with SID
     are shown in Fig.1 for $ H = 1500$ km. As expected, the paths
     are approximately U-shaped and the travel times are about $10^{5}$
     s for initial separation $x_{0} \approx 18$ m and depend on $H$
     and initial particle position. The U-shaped paths exist in a
     narrow range of "impact parameters" $y_{0}$ connected with the
     natural length scale along the $y$ axis, the separation $\Delta$
     between the libration points $L_{1}$ and $L_{2}$ (unstable
     equilibrium points situated "over" and "under" the shepherd):
\begin{equation}
       |y_{0}| \leq \Delta \approx 2a[G(M+m)/(3GM_{E})]^{1/3}.  
\end{equation}

          The trajectories are slightly asymmetric: the lower
     half is nearly straight while the upper one contains a
     significant sinusoidal component with the shepherd's orbital
     period and the amplitudes $a_{\sin}$ depending on $H$ and the
     initial data. Thus, for $H = 1500$ km, $x_{0} = 18$ m, $y_{0} =
     -25$ cm the amplitude $a_{\sin}$ is about 2 mm. The $x_{0}$ and
     $y_{0}$ dependence of $a_{\sin}$ shows that the origin of the
     oscillatory component can be connected with the nature of SID as
     "switching on" the shepherd-particle interaction at the starting
     position. Such a "cutoff" should result in a path different from
     a perfect horseshoe orbit near its turning point.

          Simulations with different initial velocities $v_{x0}$
     confirm this conclusion: for $v_{x0}$ faintly different from
     SID the value $a_{\sin}$ varies. The oscillations can occur at
     one or both branches of the trajectory; for $v_{x0}$ smaller
     than at SID they exist only at the lower branch. Larger
     deflections from SID lead to larger $a_{\sin}$; for sufficiently
     large $|v_{x0}|$ the trajectories contain loops (Fig.2).

          The initial velocity range providing a sufficiently long
     particle travel within the capsule, is rather narrow and depends
     on $H$ and $y_{0}$. In particular, for $H = 1500$ km and $y_{0} =
     -25$ cm the allowed initial velocity values are

      $v_{y0} < 0.025$ cm/s,  -0.0425 cm/s $< v_{x0} < -0.028 $ cm/s.

          The trajectories proved to be stable under variations of the
     initial position $(x_{0}, y_{0})$.

\vskip 1.0ex
          5. Trajectory dependences on the values of G (the product
     $GM_E$, known with a good accuracy, remaining invariable), the
     5th force parameter $\alpha$ for $\lambda$ of the order of
     meters, and possible EP violation ($\eta$) have been studied. As
     the variations $\delta x(t)$ turned out to be significantly
     greater than $\delta y(t)$, we speak only of $\delta x$. The
     main results are:

\begin{description}
\item[(a)]
     $\delta x(\delta G)$ and $\delta x(\delta\alpha)$ grow
     with growing $H$: they are approximately doubled when $H = 1500$
     km is changed for $H = 3000$ km.
\item[(b)]
      $y_{0}$-dependence: $\delta x(\delta G)$ and $\delta x
     (\delta\alpha)$ are the greatest for $y_{0} \approx -(1/3) \Delta
     (\approx -18$ cm for $H = 3000$ km).
\item[(c)]
      For U-shaped trajectories $\delta x(\delta G)$ is the
     greatest near the turning point.
\item[(d)]
     For looped trajectories the maximum values of $\delta
     x(\delta G)$ are about an order of magnitude greater and those of
     $\delta x (\delta\alpha)$ are nearly tripled as compared with the
     U-shaped paths; the dependence $\delta x (\eta)$ remains practically
     the same. Thus in general the looped trajectories are more promising
     from the experimental viewpoint.
\item[(e)]
     Numerically, the maximum variations $\delta x$ are:

       $ \sim 10^{-3}$ cm for $\delta G/G \sim 10^{-6}$,

 $ \sim 5 \cdot 10^{-3}$ cm for $\delta\alpha\sim10^{-5}\:(\lambda\sim 1$ m),

       $ \sim 2 \cdot 10^{-3}$ cm for  $\eta  \sim 10^{-14}$.

     These estimates confirm the viability of the proposed experiment.
\item[(f)]
     The variations $\delta x$ behave both qualitatively and
     quantitatively different at different parts of the trajectories
     under variations of $G, \alpha $ and $\eta$, allowing one to hope
     that these effects can be separated in an actual experiment.
\end{description}

\vskip 1.0ex
          6. It has been found that the quadrupole component of the
     Earth's potential causes a common displacement of the
     trajectories within about 12 cm (for $y_{0} = -25$ cm and $H =
     1500$ km) while all the effects connected with $G, \alpha $ and
     $\eta$ variations remain practically the same as those with the
     purely spherical potential.

\vskip 1.0ex
          7. The above basic features of particle motion are preserved
     when the shepherd moves along elliptic orbits with small
     eccentricities $e$ but some new features appear.

          With nonzero $e$ the sinusoidal component of particle
     trajectories becomes unavoidable and $a_{\sin}$ grows with
     growing $e$; when $e > 0.01$, loops inevitably appear. As before,
     $a_{\sin}$ grows when $v_{x0}$ deflects from SID: loops either
     appear or increase in number (Fig.3).

          Unlike the circular orbit case, the loops become tilted and
     (of possible interest for an actual experiment) increased
     $v_{x0}$ lead to trajectory squeezing in the $y$ direction,
     providing its confinement inside the capsule and creating a hope
     to use orbits with high eccentricities. However, simultaneously
     the turning points of the trajectories become remoter from the
     shepherd (Fig.4). The sensitivity of trajectories under
     gravitational interaction parameter variations are practically
     the same as that for circular orbits of the shepherd.

\vskip 1.0ex
          8. Among the possible sources of error, we examined shepherd
     nonsphericity and inhomogeneity by using multipole expansions of
     its gravitational field. We concluded that for a measurement of
     $G$ up to 1 ppm the shepherd nonsphericity $\delta R_{0}/R_{0}\,
     (R_{0}$ being its radius) should not exceed 80 ppm, or about
     $1.6\cdot 10^{-3}$ cm. Large-scale density inhomogeneities (of the order
     of $R_{0}$) must be within $1.5\cdot 10^{-3}$ and small-scale ones
     (smaller that $R_{0}/10$) within 0.07. All these requirements are
     easily met by modern technology.

\vskip 1.0ex
          9. Particle trajectory measurements are carried out with
     respect to capsule walls where the instruments are placed. The
     capsule and other bodies are sources of many sorts of noise,
     including fundamentally unavoidable, like thermal ones, which
     thus restrict the measurement accuracy. We considered the
     following basic sources of thermal noise:

          (a) radial oscillations of the shepherd's surface;

          (b) longitudinal oscillations of the capsule;

          (c) transversal oscillations of the capsule.

          Spectral analysis of thermal noises with the aid of the
     fluctuation-dissipation theorem [8] has shown that the maximum
     noise-induced measurement error does not exceed $2.5\cdot
     10^{-12}$ cm, much smaller than the expected measurement error.

\vskip 1.0ex
          10. The gravitational constant measurement procedure was
     modelled for U-shaped trajectories by three methods of $G$
     determination with the aid of
     Eqs.(1, 2): (i) the differential method, directly using the
     equations, (ii) the two-point method, employing their first
     integral, and (iii) the integral method, comparing an empirical
     trajectory with a calculated one and fitting them by varying $G$.

          The first method has the advantage of measuring $G$ at any
     small part of the trajectory, irrespective of the initial data,
     to obtain a large set of independent estimates and to use
     averaging methods to improve their accuracy. Its shortages are
     connected with relatively low accuracies with which accelerations
     and velocities can be determined. Thus, if lengths are measured
     up to $10^{-6}$ cm, the error is $\delta G/G \sim 3\cdot 10^{-5}$.

\vskip 1.0ex
          11. The two-point method employs the first integral of
     Eqs.(1, 2)
\begin{equation}
     \dot{x}^{2} + \dot{y}^{2} - \frac{2G}{s}(M+m) - 3\omega^{2}y^{2} +
          \frac{\omega^{2}}{a}y(2y^{2}-3x^{2}) = {\rm const}.
\end{equation}
          The constant $G$ is estimated by two points with known
     coordinate and velocity values, for instance, the starting and
     turning points. In the latter the velocity $v$ and the coordinate $y$
     are zero, thus removing two sources of error.

          An analysis shows that $G$ is best of all found from a set
     of independent estimates in the vicinity of the turning point.
     The achievable accuracy at the best trajectories (those with the
     turning point at 1.55-2.35 m from the shepherd) is to $\delta G/G
     \approx 4\cdot 10^{-6}$ if lenghts are measured up to $\delta l
     \sim 10^{-6}$ cm.

\vskip 1.0ex
          12. In the integral method, the most powerful one, $G$ is
     evaluated from the minimum of the functional
\begin{equation}
          S(G) = \sum_{k=1}^{n}[(x_{k}^{\rm e} - x_{k})^{2} +
                                 (y_{k}^{\rm e} - y_{k})^{2}]
\end{equation}
     measuring a ``distance'' between the two trajectories: the
     calculated one, $\{x(t), y(t)\}$, with a prescribed value of $G$
     taken for true, and an "empirical" one, $\{x^{\rm e}(t), y^{\rm e}(t)\}$,
     with a Gaussian noise corresponding to the measurement error
     $\delta l$ inserted at all ``observation'' points separated by
     equal time intervals $\Delta t$. This enabled us to estimate the
     bias ($6 \cdot 10^{-9}$) and random ($4 \cdot 10^{-8}$) errors
     $\delta G/G$ (at best) for $\delta l = 10^{-6}$ cm.

          At the present stage of the study the achievable $G$
     determination accuracy by the integral method can be estimated as
     $\delta G/G \sim 10^{-7}$ for $\delta l \sim 10^{-6}$ cm and
     $\delta G/G \sim 10^{-6}$ for $\delta l \sim 10^{-4}$ cm.

          The latter estimate is of particular significance due to the
     orbit uncertainty effect (see Table 1): evidently one can measure
     $G$ within $10^{-6}$ by tracking either small segments of
     particle trajectories for times $\sim 1$ hour with $\delta l \sim
     10^{-6}$ cm, or larger segments for times $ \sim 10$ hours with
     $\delta l \sim 10^{-4}$ cm.

          Both the two-point and integral methods admit improvements
     of the experimental data processing algorithms. In particular, in
     the integral method the bias error can be in principle entirely
     eliminated.

\vskip 1.0ex
          A general conclusion is that the SEE experiment, if
     realized, can improve our present knowledge of $G, \alpha$ (for
     certain $\lambda$) and $\eta$ by at least two orders of magnitude.

          More details are presented in a series of papers submitted to
     Izmeritelnaya Tekhnika (Russia) [9]. An alternative class of
     particle trajectories (elliptic and hyperbolic ones near the
     libration points over and under the shepherd) is analyzed in
     Ref.[10].
\vskip5mm
\centerline{\bf Acknowledgment}
\vskip5mm
      The authors are sincerely grateful to A.Sanders for
      fruitful discussions.
\pagebreak

{}
\end{document}